\documentclass[journal]{IEEEtran}
\IEEEoverridecommandlockouts
\usepackage{setspace}
\usepackage{cite}
\usepackage{amsmath,amssymb,amsfonts}
\usepackage{graphicx}
\usepackage{textcomp}
\usepackage{xcolor}
\usepackage{algpseudocode}
\usepackage{mathtools} 
\usepackage{amsmath}
\usepackage{amssymb}
\usepackage{amsthm}
\usepackage{mathtools}
\usepackage[ruled, vlined, linesnumbered]{algorithm2e}
\usepackage{color,soul}

\usepackage{cite}
\usepackage{subfigure}
\usepackage{stmaryrd}
\usepackage{breqn}
\usepackage{stackengine}
\usepackage{verbatim} 
\usepackage{glossaries}
\usepackage{color,soul}
\usepackage{xcolor}
\def\BibTeX{{\rm B\kern-.05em{\sc i\kern-.025em b}\kern-.08em
    T\kern-.1667em\lower.7ex\hbox{E}\kern-.125emX}}
\begin{document}

\title{\huge Meta-Learning-Driven Resource Optimization in Full-Duplex ISAC with Movable Antennas\\
}
\author{Ali Amhaz, Shreya Khisa, Mohamed Elhattab, Chadi Assi and Sanaa Sharafeddine \vspace{-0.6cm}}
%

\maketitle
\begin{abstract} 
This paper investigates a full-duplex (FD) scenario where a base station (BS) equipped with movable antennas (MAs) simultaneously provides communication services to a set of downlink (DL) and uplink (UL) users while also enabling sensing functionalities for target detection, thereby supporting integrated sensing and communication (ISAC) technology. Additionally, a receiving BS, also equipped with MAs (denoted as BS R), is responsible for capturing the reflected echo. To optimize this setup, we formulate an optimization problem aimed at maximizing the signal-to-noise and interference ratio (SINR) of the captured echo. This is achieved by jointly optimizing the transmit beamforming vectors at the FD BS, the receiving beamforming vectors at both the FD BS and BS R, the UL users' transmit power, and the MAs' positions at both BSs, all while satisfying the quality-of-service (QoS) requirements for both sensing and communication. Given the non-convex nature of the problem and the high coupling between the variables, we employ a gradient-based meta-learning (GML) approach tailored for large-scale optimization. Numerical results demonstrate the effectiveness of the proposed meta-learning approach, achieving results within 99\% of the optimal solution. Furthermore, the MA-based scheme outperforms several benchmark approaches, highlighting its advantages in practical ISAC applications.
\end{abstract}
 
\begin{IEEEkeywords}
Beamforming, Movable Antenna , Full-duplex, meta-learning, ISAC.
\end{IEEEkeywords}
\vspace{-0.3cm}
\section{Introduction}
Integrated sensing and communication (ISAC) has emerged as one of the enabler for sixth-generation (6G) networks, aiming to unify sensing and communication functionalities within a shared infrastructure. By jointly utilizing spectral and hardware resources, ISAC significantly improves spectrum efficiency and enables a wide array of emerging applications \cite{9737357}. Complementing ISAC, full-duplex (FD) communication has gained increasing attention, particularly at base stations (BSs), due to its ability to transmit and receive simultaneously over the same frequency band. This simultaneous operation has the potential to nearly double spectral efficiency (SE) by supporting concurrent downlink and uplink transmissions, positioning FD as a promising solution to meet the stringent performance and spectrum demands of future 6G networks. Moreover, some studies demonstrated the promising advantages of integrating both technologies \cite{10373185}.

Although FD technology and ISAC  offer substantial benefits for future wireless networks, they also present significant challenges that may impede their practical implementation and performance gains. FD technology suffers from residual self-interference (SI) due to the transmitting antennas affecting the receiving side, as well as interference from uplink (UL) transmissions impacting downlink (DL) users. Meanwhile, ISAC systems face complex trade-offs and design challenges, as the simultaneous provision of sensing and communication services can lead to conflicting requirements in some cases.

Movable antennas (MAs) have recently caught attention due to their ability to be flexibly positioned, mitigating interference, battling severe fading scenarios, and yielding higher SE and multiplexing gains as demonstrated in \cite{10318061} \cite{khisa2025metalearningdrivenmovableantennaassistedfullduplex}. 
The authors in \cite{10318061} examined an MA-enabled multiple-input multiple-output (MIMO) system to maximize the sum rate by optimizing MA positions, demonstrating significant improvements over fixed-position antenna (FPA) systems. In \cite{10839251}, an ISAC system leveraging MA technology was analyzed to jointly maximize the communication rate and sensing function. To address challenges associated with FD technology, \cite{khisa2025metalearningdrivenmovableantennaassistedfullduplex} proposed an MA-enabled FD system aimed at maximizing the achievable sum rate for both UL and DL users. The related literature remains scarce, with the notable exception of \cite{guo2024movableantennaenhancednetworked}, which investigated an MA-aided FD system supporting ISAC functionality. While this work thoroughly examined the concept, it assumed separate transmitting and receiving BSs for communication, thereby increasing system complexity and neglecting the interference from transmitting BSs to receiving BSs. Furthermore, the adopted successive convex approximation (SCA) approach may not scale efficiently with a large number of decision parameters.

\par Motivated by the aforementioned studies, we investigate, in this letter, an FD ISAC system enhanced with MAs to mitigate interference, improve channel conditions, and enable simultaneous communication and sensing. Through dynamic repositioning, MAs provide the flexibility needed to effectively manage the inherent trade-offs in ISAC systems while fully exploiting the benefits of FD architecture. 
In this regard, we formulate an optimization problem intending to maximize the signal-to-noise and interference ratio (SINR) of the reflected echo while meeting the quality-of-service requirements (QoS) for both sensing and communication. Due to the non-convex structure of the formulated optimization problem, we adopt a gradient-based meta-learning (GML) approach tailored for large-scale optimization. 

  \vspace{-0.3cm}
\section{System Model}
\begin{figure}[!t]
    \centering    \includegraphics[width=1\columnwidth]{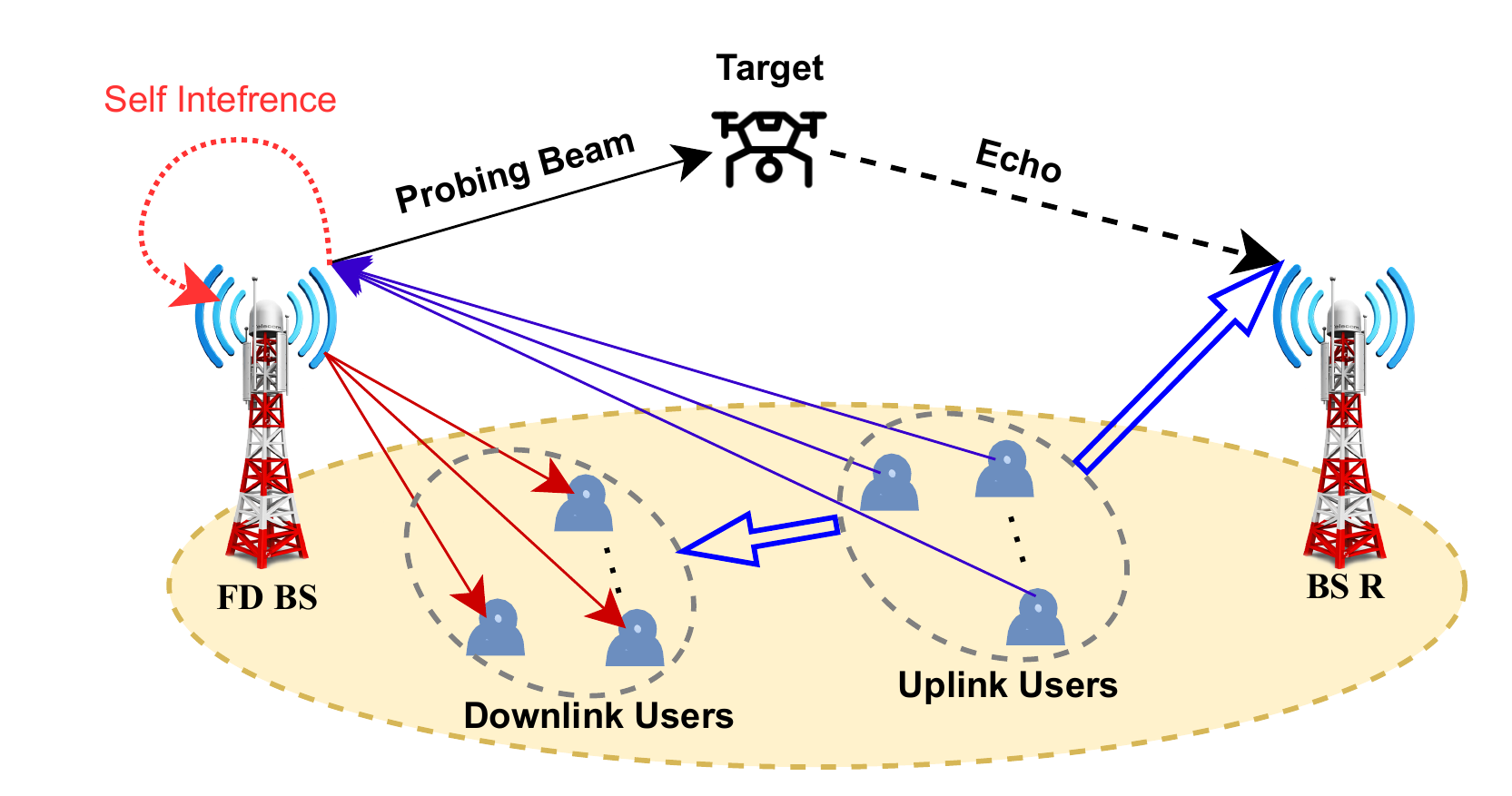}
    \caption{System model.}
	\label{fig_11}
    \vspace{-0.5cm}
\end{figure}

The system model in this work comprises an FD BS, equipped with $N_T$ transmitting MAs, with the coordinates $\textbf{t}^{BS}_{n_t}=[x_{n_t},y_{n_t}]^T (1 \le n_t\le N_T)$, and $N_{R_c}$ receiving MAs denoted as $\textbf{r}^{BS}_{n_{r_c}}=[x_{n_{r_c}},y_{n_{r_c}}]^T   (1 \le n_{r_c}\le N_{R_c})$, responsible for serving a set of $D$ DL users and $U$ UL users, as depicted in Fig. \ref{fig_11}. The DL users with FPA location $\textbf{r}^{d}_{i}=[x_{i},y_{i}]^T \forall i \in \mathcal{D}$ are indexed by the set $\mathcal{D} = \{1,..,D\}$, while the UL users with location $\textbf{t}^{u}_{j}=[x_{j},y_{j}]^T \forall j \in \mathcal{U}$ are represented by the set $\mathcal{U} = \{1,..,U\}$. Now, due to the FD nature of the serving BS, a residual SI will affect the received UL transmission. On the other hand, by empowering ISAC technology, the FD BS will generate a probing beam to provide sensing functionality for a target $t$. After hitting $t$, the signal will bounce back to a receiving BS denoted as $R$ and equipped with $N_{R_s}$ MAs with the coordinates $\textbf{r}^{BS}_{n_{r_s}}=[x_{n_{r_s}},y_{n_{r_s}}]^T   (1 \le n_{r_s}\le N_{R_s})$. This bi-static architecture demonstrated resilience against fast-reflecting echo signals that lead to interference problems \cite{10207991}.
%
\vspace{-0.4cm}
\subsection{The Channel Model: Communication Functionality}
Following the approach in \cite{10318061}, we adopt the field response channel model for all communication links in the proposed system. 
The DL channel between the FD BS and the DL user $i$ is defined as $\boldsymbol{h}_{d, i}$ while the channel between the UL user $j$ and the FD BS is expressed as $\boldsymbol{h}_{u,j}$. On the other hand, the channel between the UL user $j$ and the DL user $i$ can be expressed as $h_{j,i}$. The direct interference channel between
the UL user $j$ and BS R is $\boldsymbol{h}_{j,r}$. Also, the SI channel is defined as $\boldsymbol{H}_{SI}\in \mathbb{C}^{N_T \times N_{R_c}}$. In the FD BS, for the transmitting side, the origin of the antenna $n_t$ is indicated as $O_{n_t}$, while in the receiving component, the origin of the antenna $n_{r_c}$ is expressed as $O_{n_{r_c}}$. From the FD BS to the DL user $i$, we express the number of transmit paths as $L_{i,d}^T$, while the number of receive paths for the same user can be expressed as $L_{i,d}^R$. Similarly, we define the number of transmit and receive paths between the FD BS and the UL user $j$ as $L_{j,u}^T$ and $L_{j,u}^R$, respectively. Now, the path response matrix (PRM) related to the channel between the FD BS and the DL user $i$ is expressed as $\boldsymbol{\Sigma}_{d, i} \in \mathbb{C}^{L_{i,d}^R \times L_{i,d}^T}$. Also, the PRM between the UL user $j$ and the FD BS is expressed as $\boldsymbol{\Sigma}_{u, j} \in \mathbb{C}^{L_{j,u}^R \times L_{j, u}^T}$. Following the same strategy, the PRM corresponding to the interference link between the UL user $j$ and the DL user $i$ is expressed as $\boldsymbol{\Sigma}_{i, j} \in \mathbb{C}^{L_{i,j}^R \times L_{j,u}^T}$ \cite{zhu2024movableantennaenhancedmultiusercommunication}. 
The divergence in the signal propagation for the $m$-th transmit path of the FD BS's $n_t^{th}$ transmit antenna (acquiring the location  $\textbf{t}^{BS}_{n_t}=[x_{n_t},y_{n_t}]^T$) and its origin can be represented as $\rho^t_{m,i} = x_{n_t}^t \cos \theta^t_{m,i} \sin \phi^t_{m, i} + y_{n_t} \sin \theta_{m,i} $. Here, $\theta_{m,i}$ and  $\phi_{m,i}$ correspond to the elevation and azimuth of the angle of departure (AoD) \cite{10318061}. In light of this, the transmit field response vector (FRV) of the $n_t^{th}$ antenna at the BS is defined as:
\begin{align}
&\textbf{g}_i\left(\boldsymbol{t}_{n_t}^{BS}\right) = \notag \\ &[e^{-j \frac{2\pi}{\lambda} \rho^t_{1, i} (\boldsymbol{t}_{n_t}^{BS})}, e^{-j \frac{2\pi}{\lambda} \rho^t_{2, i} (\boldsymbol{t}_{n_t}^{BS})}, ..., e^{-j \frac{2\pi}{\lambda} \rho^t_{L_{i,d}^T, i} (\boldsymbol{t}_{n_t}^{BS})} ],
\end{align}
%
\noindent Similarly, the FRV received by the DL user $i$ is derived as: 
\begin{align}
    &\textbf{f}_i^d = [e^{-j \frac{2\pi}{\lambda} \rho^r_{1, i} (\boldsymbol{r}_i^d)}, e^{-j \frac{2\pi}{\lambda} \rho^r_{2, i} (\boldsymbol{r}_i^d)}, ..., e^{-j \frac{2\pi}{\lambda} \rho^r_{L_{i,d}^R, i} (\boldsymbol{r}_i^d)} ],
\end{align}
\noindent where the corresponding propagation difference of the $q$-th receiving path between the location of the antenna and its origin is defined as $\rho^r_{q,i} = x_i^r \cos \theta^r_{q,i} \sin \phi^r_{q, i} + y_i^r \sin \theta_{q,i}$, with $\theta^r_{q,i}$ being the elevation angle of arrival (AoA) and $\phi^r_{q, i}$ the azimuth. Thus, the channel $\boldsymbol{h}_{d, i}$ can be derived as:
\begin{align}
    &\boldsymbol{h}_{d, i} = (\boldsymbol{\text{f}_i^d})^H. \boldsymbol{\Sigma}_{d, i} . \boldsymbol{G}_i, 
\end{align}
\noindent where $\boldsymbol{G}_i = [\textbf{g}_i(\boldsymbol{t}_1^{BS}), \textbf{g}_i(\boldsymbol{t}_2^{BS}), ..., \textbf{g}_i(\boldsymbol{t}_{N_T}^{BS}) ] $ corresponds to the transmit field response matrix (FRM) from the FD BS to the DL user $i$. Adopting the same procedure, we derive the channel   $\boldsymbol{h}_{u, j} = \boldsymbol{\text{F}}_j^H (\boldsymbol{r}_j^u) . \boldsymbol{\Sigma}_{u, j} . \boldsymbol{g}_j,$
where $\boldsymbol{F}_j (\boldsymbol{r}_j^u) = [\textbf{f}_j(\boldsymbol{r}_1^{u}), \textbf{f}_j(\boldsymbol{r}_2^{u}), ..., \textbf{f}_j(\boldsymbol{r}_{N_{R_c}}^{u}) ]$ is the corresponding FRM from the UL user $j$ to the FD BS, with 
\begin{align}
    &\textbf{f}_j (\boldsymbol{r^u_{n_{r_c}}}) = [e^{-j \frac{2\pi}{\lambda} \rho^r_{1, j} (\boldsymbol{r}_i^d)}, e^{-j \frac{2\pi}{\lambda} \rho^r_{2, j} (\boldsymbol{r}_i^d)}, ..., e^{-j \frac{2\pi}{\lambda} \rho^r_{L_{j,u}^R, j} (\boldsymbol{r}_i^d)} ],
\end{align}
where the propagation difference $\rho^r_{q,j}$ is obtained similarly as  before while $\textbf{g}_j$ is defined as
\begin{align}
    &\textbf{g}_j = [e^{-j \frac{2\pi}{\lambda} \rho^t_{1, i} (\boldsymbol{t}_j^{u})}, e^{-j \frac{2\pi}{\lambda} \rho^t_{2, i} (\boldsymbol{t}_j^{u})}, ..., e^{-j \frac{2\pi}{\lambda} \rho^t_{L_{j,u}^T, i} (\boldsymbol{t}_j^{u})} ].
\end{align}
Now, for the channel $\boldsymbol{h}_{j, i}$ between the DL user $i$ and UL user $j$ it is defined as $\boldsymbol{h}_{j, i} = \boldsymbol{\text{f}}_{i,j}^H . \boldsymbol{\Sigma}_{i, j} . \boldsymbol{g}_{i,j}, $
%
where $\boldsymbol{\text{f}}_{i,j}$ and $\boldsymbol{g}_{i,j}$ are the corresponding receive and transmit FRV. For channel $\boldsymbol{H}_{SI}$\cite{khisa2025metalearningdrivenmovableantennaassistedfullduplex}, it is assumed that the number of transmit and receive paths are respectively, $L_{SI}^T$ and $L_{SI}^R$, with PRM  $\boldsymbol{\Sigma}_{SI}\in \mathbb{C}^{L_r^{SI}\times L_t^{SI}}$. Thus, it is  expressed as $\textbf{H}_{SI}=(\textbf{F}_{SI}^H(\textbf{r}_{n_{r_c}}^{BS})\boldsymbol{\Sigma}_{SI}\textbf{G}_{SI}(\textbf{t}_{n_t}^{BS}))^T$, 
where 
\begin{equation}
    \textbf{F}_{SI}(\textbf{r}_{n_{r_c}}^{BS})=[\textbf{f}_{SI}(\textbf{r}_1^{BS}),\dots,\textbf{f}_{SI}(\textbf{r}_{N_{R_c}}^{BS})] \in \mathbb{C}^{L_r^{SI}\times N_{R_c}},
\end{equation}
\begin{equation}
\textbf{f}_{SI}(\textbf{r}_{n_{r_c}}^{BS})=[e^{j\frac{2\pi}{\lambda}\rho_{r,1}(\textbf{r}_{n_r}^{BS}
    )},\dots,e^{j\frac{2\pi}{\lambda}\rho_{r,L^{SI}_{r}}(\textbf{r}_{n_r}
    )}]^T \in \mathbb{C}^{L_r^{SI}\times 1},
    \end{equation} 
\begin{equation}
    \textbf{G}_{SI}(\textbf{t}_{n_t}^{BS})=[\textbf{g}_{SI}(\boldsymbol{t}_{1}^{BS}),\dots,\textbf{g}_{SI}(\boldsymbol{t}_{N_T}^{BS})] \in \mathbb{C}^{L_t^{SI} \times N_T},
    \end{equation}
    \begin{equation}
\textbf{g}_{SI}(\textbf{t}_{n_t}^{BS})=[e^{j\frac{2\pi}{\lambda}\rho_{n_t,1}(\textbf{t}_{nt}^{BS}
    )},\dots,e^{j\frac{2\pi}{\lambda}\rho_{n_t,L^{SI}_{t}}(\textbf{t}_{n_t}^{BS}
    )}]^T \in \mathbb{C}^{L_{t}^{SI} \times 1},
    \end{equation}
    %
where $\textbf{F}_{SI}(\textbf{r}_{n_{r_c}}^{BS})$ and $\textbf{G}_{SI}(\textbf{t}_{n_t}^{BS})$ are the receive and transmit FRMs, respectively. Finally, the direct interference link between the UL user $j$ and BS R is expressed as $\boldsymbol{h}_{j,r}$ where it follows a similar model to that of the UL communication channel $\boldsymbol{h}_{u,j}$ and the full details are eliminated for brevity.
\vspace{-0.3cm}
\subsection{The Channel Model: Sensing Functionality}
The channel between the FD BS and the target $t$ can be expressed as
\begin{equation}
    \boldsymbol{g}_{b,t} = \beta_{b,t} \hat{\boldsymbol{g}_{b,t}},
\end{equation}
where $\beta_{b,t}$ is a complex fading coefficients related to the target and $\hat{\boldsymbol{g}}_{b,t}$ is a vector defined as:
\begin{align}
&\hat{\boldsymbol{g}}_{b,t} = [e^{-j \frac{2\pi}{\lambda} \rho^t_{1, t} (\boldsymbol{t}_n^{BS})}, e^{-j \frac{2\pi}{\lambda} \rho^t_{2, t} (\boldsymbol{t}_n^{BS})}, ..., e^{-j \frac{2\pi}{\lambda} \rho^t_{L_{i,d}^T, t} (\boldsymbol{t}_n^{BS})} ],
\label{eee}
\end{align}
where $\rho^t_{m,t} = x_{n_t}^t \cos \theta^t_{m,t} \sin \phi^t_{m, t} + y_{n_t} \sin \theta_{m,t}$ with $\theta^t_{m,t}$ and $\phi^t_{m, t}$ being the elevation and azimuth angle to the target, respectively.
Similarly, the channel between the target and BS R is defined as $\boldsymbol{g}_{sr,t} = \beta_{sr,t} \hat{\boldsymbol{g}}_{sr,t}$ where $\beta_{sr,t} $ is a complex fading coefficient and $\hat{\boldsymbol{g}}_{sr,t}$ is expressed similar to \eqref{eee}. 
  \vspace{-0.3cm}
\subsection{Signal Model}
The signal transmitted by the FD BS to serve the communication users and perform sensing is expressed as $\textbf{x}= \sum_{i \in {D}}\textbf{p}_{d,i} s_{d,i},$
%
\noindent where $s_{d,i}$ is the signal intended for the DL user $i$ and $\boldsymbol{p}_{d,i}$ being the corresponding beamforming vector. Meanwhile, the signal transmitted by the UL user $j$ to the FD BS is defined as $x_{u,j} = \sqrt{P_{u,j}}s_{u,j}, \forall j \in \mathcal{U}$,
%
\noindent where $s_{u,j}$ is the signal transmitted and $P_{u,j}$ representing the transmit power. Now, the signal received by the DL users is defined as
\begin{align}
y_{d,i} = 
\textbf{h}_{d,i}^H(\textbf{p}_{d,i}s_{d, i}) & +
\sum_{i' \in \mathcal{D}, i' \neq i}\textbf{h}_{d,i}^H\textbf{p}_{d, i'}s_{d, i'}
+
\sum_{j \in \mathcal{U}}h_{j,i}x_{u,j}  \notag \\ &+n_{d,i}, 
\label{down_rec}
\end{align}
where the first term corresponding to the desired signal, the second term to the interference coming from the signal intended to other users, and the third one representing the UL interference, and $n_{d, i}$ being the additive white Gaussian noise (AWGN) following $\mathcal{CN}(0, \sigma_{d,i}^2)$. Correspondingly, the rate to decode the signal of DL user $i$ is presented as:
\begin{align}
&R_{d, i} = \log_2\left(1+ \right. \notag \\ & \left. \frac{|\textbf{h}_{d,i}^H\textbf{p}_{d,i}|^2}{\sum_{i' \in \mathcal{D}, i' \neq i}|\textbf{h}_{d,i}^H\textbf{p}_{d, i'}|^2+
    \sum_{j \in \mathcal{U}}P_{u,j}|h_{j,i}|^2 + \sigma_{d,i}^2}\right).
\label{d_r}
\end{align}
Now, the signal received at the FD BS due to the UL transmission can be expressed as follows:
\begin{equation}
{\textbf{y}_c}=\sum_{j \in \mathcal{U}}\textbf{h}_{u,j}x_{u, j}+\textbf{H}_{SI}{\textbf{x}}+\textbf{n}_{BS}^u,
\end{equation}
 where $\textbf{n}_{BS}^u \sim \mathcal{CN}(0,\sigma_u^2\textbf{I}_{N_R})\in \mathbb{C}^{N_R \times 1}$ is the AWGN  with zero mean and variance $\sigma_u^2\textbf{I}_{N_R})$. In light of this, the rate to decode the signal of user $j$ is represented as:
  \begin{equation}
R_{u,j}=\log_2\left(1+ \frac{P_{u,j}|\textbf{Z}_{u,j}^H\textbf{h}_{u,j}|^2}{{I_{UI}+
I_{SI}+
\sigma_u^2||\textbf{Z}_{u,j}^H||_2^2}}\right),
\label{u_r}
 \end{equation}
where $I_{UI} = \sum_{j' \in \mathcal{U}, j' \neq j}P_{u,j'}|\textbf{Z}_{u,j}^H\textbf{h}_{u,j'}
|^2$, $I_{SI} = |\textbf{Z}_{u,j}^H|^2|\textbf{H}_{SI}^H 
(\sum_{i \in \mathcal{D}}\textbf{p}_{d,i})|^2$, $\boldsymbol{Z}_{u,j}$ is the receive beamforming vector intended for UL user $j$ and $||.||_2$ is Euclidean norm.
\vspace{-0.3cm}
\subsection{The Sensing Model}
At the receiving BS R, the signal of the echo is defined as:
\begin{align}
{\textbf{y}_s} = 
\sum_{i \in \mathcal{D}}\boldsymbol{\alpha}_t  \textbf{g}_{sr, t}   \textbf{g}_{b, t}^H \boldsymbol{x} &+
\sum_{c \in \mathcal{C}}\sum_{i \in \mathcal{D}}\boldsymbol{\alpha}_c  \textbf{g}_{sr, c}   \textbf{g}_{b, c}^H \boldsymbol{x}
+ \sum_{j \in \mathcal{U}}\textbf{h}_{j,r}
x_{u,j} \notag \\ &+
\textbf{n}_{BS}^s, 
\label{sesning}
\end{align}
\noindent where $\boldsymbol{\alpha}_t$ and $\boldsymbol{\alpha}_c$ correspond to the target and clutter radar cross section (RCS). The first term in \eqref{sesning} represents the useful echo signal reflected from the target $t$ and captured by the receiving BS R, while the second term accounts for the echo signal originating from clutter. The third term corresponds to the direct UL transmission from the UL users to BS R. Finally, $\boldsymbol{n}_{BS}^s$ denotes the AWGN. Based on the above expression, we adopt the SINR of the received echo at BS R as the key performance metric, and it is expressed as:
\begin{align}
\hat{\Lambda}_t=  
\frac{\sum_{i \in \mathcal{D}}\boldsymbol{\alpha}_t|\textbf{V}^H \textbf{g}_{sr, t}   \textbf{g}_{b, t}^H \textbf{p}_{d,i}|^2}{I_D + I_C +\sigma_u^2||\textbf{r}_{t}^H||^2_2},
\label{SCINR_sesning}
\end{align}
where $\boldsymbol{V}$ is the receiving beamforming vector at BS R, $I_D = \sum_{j \in \mathcal{U}}P_{u,j}|\textbf{V}^H\textbf{h}_{j,r}
|^2$, and $I_c = \sum_{c \in \mathcal{C}} \sum_{i \in \mathcal{D}}\boldsymbol{\alpha}_c|\textbf{V}^H \textbf{g}_{sr, c}   \textbf{g}_{b, c}^H \textbf{p}_{d,i}|^2$. Now, we normalize the SINR by applying the logarithmic operation \cite{10839251}. Hence, the final expression is denoted as $\Lambda_t = \log_2(\hat{\Lambda}_t)$.
  \vspace{-0.3cm}
\section{Problem Formulation}
In this work, we focus on maximizing the SINR for the received echo signal at BS $R$. In this regard, we formulate an optimization problem to maximize $\Lambda_t$ by jointly optimizing the transmit beamforming vectors at FD BS ($\boldsymbol{P} = \{ \boldsymbol{p}_{d,i}, \forall i \in \mathcal{D} \}$), the transmit power at the UL users ($\boldsymbol{P}_u = \{ P_{u,j}, \forall j \in \mathcal{U} \}$), the receive beamforming vector at FD BS ($\boldsymbol{Z} = \{ \boldsymbol{Z}_{u,j}, \forall j \in \mathcal{U} \}$), the receive beamforming vector at BS R and the positions of the MAs at the two BSs ($\boldsymbol{L} = \{\boldsymbol{t}_{n_t}^{\mathrm{BS}} : n_t \in [N_t]\} 
\cup \{\boldsymbol{r}_{n_{r_c}}^{\mathrm{BS}} : n_{r_c} \in [N_{r_c}]\} 
\cup \{\boldsymbol{r}_{n_{r_s}}^{\mathrm{BS}} : n_{r_s} \in [N_{r_s}]\}$) while meeting the QoS constraints for both communication and sensing functionalities. Accordingly, the problem can be formulated as follow:
\allowdisplaybreaks
\begingroup
\begin{subequations}
\begin{align}
&\mathcal{P}_1: \max_{\boldsymbol{P}, \boldsymbol{Z}, \boldsymbol{V}, \boldsymbol{P}_u, \boldsymbol{L}}  \ \;  \Lambda_t, \\
\textrm{s.t.}  \ \; 
     &\textrm{Tr} (\textbf{P}\textbf{P}^H) \le P_{BS}, \label{C_7} \\
   &  P_{u,j} \le P_u^{max}, \forall j \in \mathcal{U}, \label{C_2}\\
   &  R_{d,i} \ge R_{th,d},  R_{u,j} \ge R_{th,u}, \forall i \in \mathcal{D}, \forall j \in \mathcal{U}, \label{C_3}\\
   &  \Lambda_t \ge \Lambda_{th,s}, \label{C_3333}\\
   & |\textbf{Z}_{u,j}|=1, \forall j \in \mathcal{U}, \label{C_5}\\
   & |\textbf{V}|=1, \label{tx1}\\
   & \textbf{t} \in \mathcal{C}_t^{BS}, \textbf{r}_{c} \in \mathcal{C}_{r_c}^{BS},  \textbf{r}_s \in \mathcal{C}_{r_s}^{BS}\label{c4}\\
   & ||\textbf{t}_{n_t}-\textbf{t}_{\hat{n}_t}|| \ge DS, 1 \le n_t \neq \hat{n_t}\le N_T, \label{c8} \\
   & ||\textbf{r}_{n_{r_c}}-\textbf{r}_{\hat{n}_{r_c}}|| \ge DS, 1 \le n_{r_c} \neq \hat{n}_{r_c}\le N_{R_c} \label{c9},\\
   & ||\textbf{r}_{n_{r_s}}^{BS}-\textbf{r}_{\hat{n}_{r_s}}^{BS}|| \ge DS, 1 \le n_{r_s} \neq \hat{n}_{r_s}\le N_{R_s} \label{c99},
\end{align} 
\label{optim}
\end{subequations}
\endgroup
\noindent \noindent where the terms $P_{BS}$ and $P_u^{max}$ correspond to the power budget of the FD BS and the UL users, respectively. The rate thresholds of the DL and UL users are denoted as $R_{th,d}$ and $R_{th,u}$, respectively. For the sensing functionality, the SINR threshold is $\Lambda_{th,s}$.  $\mathcal{C}_{t}^{BS}$, $\mathcal{C}_{r_c}^{BS}$, and $\mathcal{C}_{r_s}^{BS}$ represent the mobility region for the transmitting antennas in FD BS, the receiving antennas in FD BS, and the receiving antennas in BS R, respectively, with $DS$ being the minimum spacing separation of the antennas. The constraints \eqref{C_7} and \eqref{C_2} are the power budget constraints at the FD BS and the UL devices, respectively. \eqref{C_3} are the QoS constraints for communication while \eqref{C_3333} is for sensing functionality. \eqref{C_5} and \eqref{tx1} constraints are for the receiving beamforming vectors at the FD BS and the BS R, respectively. The constraint \eqref{c4} is to restrict the mobility regions for the MAs at the different BSs. Finally, in order to ensure the minimum spacing between the antennas, the constraints \eqref{c8} - \eqref{c99} are imposed. 
  \vspace{-0.3cm}
\section{Solution Approach}
The optimization problem presented in the previous section proved to be non-convex and difficult to solve using the traditional optimization solvers due to the high coupling between the different variables. In this regard, we leverage a GML approach intended to solve complex, large-scale optimization problems by directly feeding the gradients $\Delta_{\boldsymbol{P}}R_{\boldsymbol{P}}$, $\Delta_{\boldsymbol{L}}R_{\boldsymbol{L}}$, $\Delta_{\boldsymbol{P}_u}R_{\boldsymbol{P}_u}, \Delta_{\boldsymbol{Z}}R_{\boldsymbol{Z}}$, and $\Delta_{\boldsymbol{V}}R_{\boldsymbol{V}}$, obtained relying on the objective function of \eqref{optim} $R_a = \Lambda_t$, $\forall a \in \{ \boldsymbol{P}, \boldsymbol{L}, \boldsymbol{P}_u, \boldsymbol{Z}, \boldsymbol{V}\}$, to a set of neural networks which will output the gradients $\Delta\boldsymbol{P}, \Delta \boldsymbol{L}, \Delta \boldsymbol{P}_u, \Delta \boldsymbol{Z},$ and $\Delta \boldsymbol{V}$ that will be added to the initialized values of the different optimization variables. This technique relies on the gradients to extract higher-order information for the optimization variables rather than the actual values, thus reducing the complexity \cite{10623434}. 
  \vspace{-0.3cm}
\subsection{Gradient-based Meta Learning (GML) Architecture}
In contrast to traditional data-driven meta-learning techniques that need an offline pre-training process,  adaptation, and online refinement, which will cause variations in the data distribution along with high energy consumption and problems for latency-critical use cases, the introduced model showed to acquire strong robustness without the need for a pre-training phase. It is built to work on the trajectory rather than the variables themselves. Our approach consists of two main components: a base learner and a meta learner. These components are established by three different layers and they can be expressed as follows:
\subsubsection{Inner Iteration}
At this level, a cyclic optimization problem is performed for the variables $\boldsymbol{P}$, $\boldsymbol{L}$, $\boldsymbol{P}_u$, $\boldsymbol{Z}$, and $\boldsymbol{V}$ in which each variable will be tackled using a sub-network:
\begin{itemize}
  \item \textbf{Precoding Network (PN)}: In this sub-network, the transmitting precoding vectors at the FD BS are optimized.  
\item \textbf{Uplink-Power Network (UPN)}: This sub-network optimizes the UL users' transmit power.  
\item \textbf{Movable Antennas Network (MAN)}: This sub-network optimizes the MAs' positions at the FD BS and BS R.
\item \textbf{Communication Receiving Beamforming Network(CRBN)}: In this sub-network, the variable $\boldsymbol{Z}$ at FD BS is optimized. 
\item \textbf{Sensing Receiving Beamforming Network(SRBN)}: This sub-network optimizes the  beamforming vector $\boldsymbol{V}$.
  \end{itemize}
All the networks defined above follow a sequential process in updating the values of the different variables, starting from pre-defined initial points. On the flip side, the values of the other four variables are obtained from their corresponding sub-networks. The update can be described as $\boldsymbol{P}_u^*=R(\boldsymbol{P}_u^{(i,j)},\boldsymbol{P}^*,$ $\boldsymbol{L}^*,\boldsymbol{Z}^*,\boldsymbol{V}^*), \boldsymbol{P}^*=R(\boldsymbol{P}^{(i,j)},\boldsymbol{P}_u^*,\boldsymbol{L}^*,\boldsymbol{Z}^*$ $,\boldsymbol{V}^*),
 \boldsymbol{L}^*=R(\boldsymbol{L}^{(i,j)} ,\boldsymbol{P}_u^*,\boldsymbol{P}^*,\boldsymbol{Z}^*,\boldsymbol{V}^*),$$
\boldsymbol{Z}^*=R(\boldsymbol{Z}^{(i,j)} ,\boldsymbol{P}_u^*, \boldsymbol{P}^*, \boldsymbol{L}^*, \boldsymbol{V}^*),$ and $
\boldsymbol{V}^*=R(\boldsymbol{V}^{(i,j)} ,\boldsymbol{P}_u^*,$ $\boldsymbol{P}^*, \boldsymbol{Z}^*, \boldsymbol{L}^*),$
\begin{figure*}
  \centering
  \subfigure[Convergence of the solution approach.]{\includegraphics[width=0.30\linewidth]{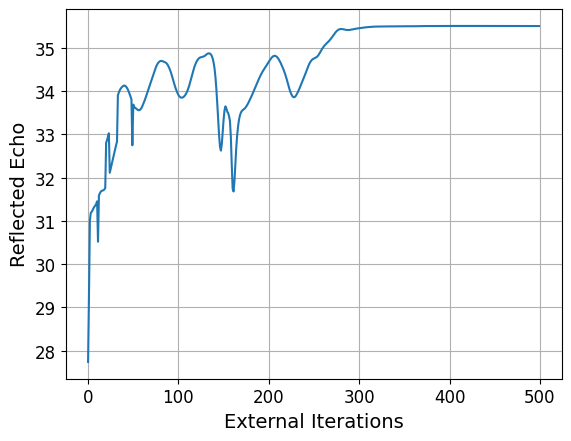}}
  \subfigure[BS power vs reflected echo.]
  {\includegraphics[width=0.34\linewidth]{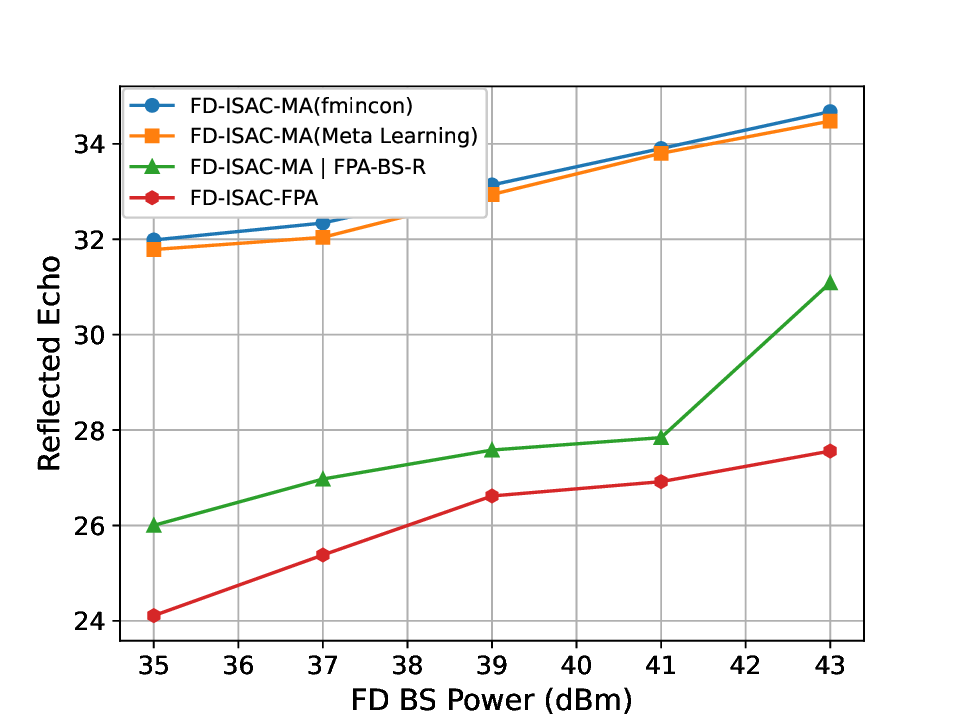}}
  \subfigure[SI vs reflected echo.]{\includegraphics[width=0.34\linewidth]{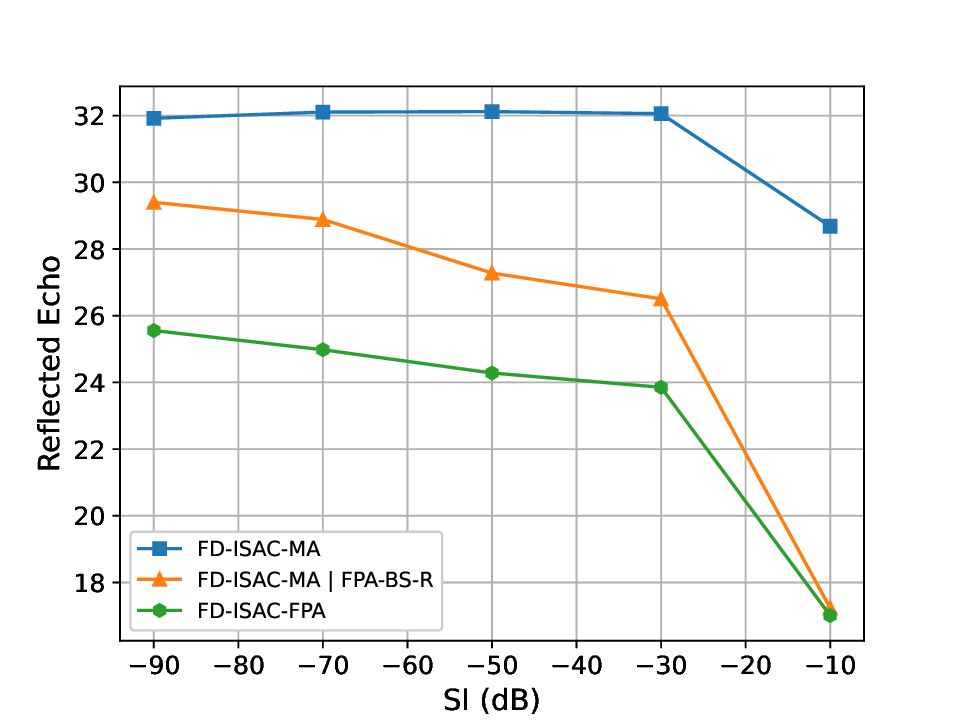}}
  \setlength{\belowcaptionskip}{-12pt} 
  \caption{Numerical results for the model under different parameters.}
  \label{three_figures}
  \vspace{-0.3cm}
\end{figure*}
\begin{figure}[!t]
    \centering    \includegraphics[width=0.67\columnwidth]{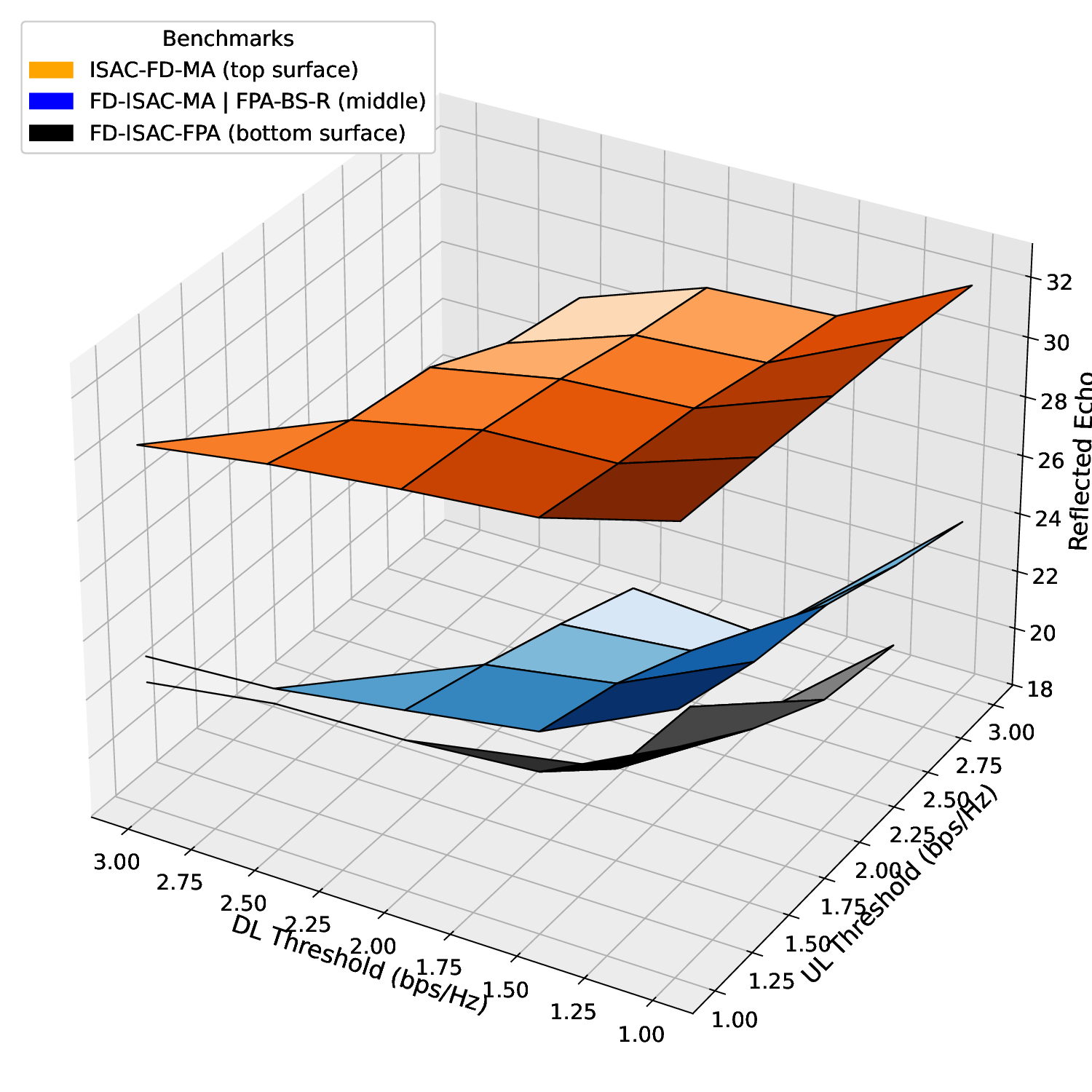}
    \caption{Reflected echo vs DL and UL rate thresholds.}
    \label{finall}
    \vspace{-0.5cm}
\end{figure}
where $\boldsymbol{P}_u^{(i,j)}$, $\boldsymbol{P}^{(i,j)}$, $\boldsymbol{L}^{(i,j)}$, $\boldsymbol{Z}^{(i,j)}$, and $\boldsymbol{V}^{(i,j)}$ are the variables values at the $j$-th outer iteration and $i$-th inner iteration. Now, we review the details of the PN network, while the other networks follow the same logic, and the full description is omitted for brevity. 

This network seeks to refine the transmit precoding vectors at the FD BS during the $j$-th outer and $i$-th inner optimization loops. The optimization objective is to maximize the overall SINR of target, expressed as $R(\boldsymbol{P}^{(i,j)},\hat{\boldsymbol{P}}_u,\hat{\boldsymbol{L}}, \hat{\boldsymbol{Z}}, \hat{\boldsymbol{V}})$, where $\hat{\boldsymbol{P}}_u,\hat{\boldsymbol{L}}, \hat{\boldsymbol{Z}},$ and $ \hat{\boldsymbol{V}}$ represent the current estimations for the variables. At each iteration, the SINR is evaluated, and the corresponding gradient with respect to $\boldsymbol{P}^{(i,j)}$ is computed. This gradient is then passed through a neural network, which outputs an adjustment term $\Delta \boldsymbol{P}^{(i,j)}$. The precoding matrix is subsequently updated as $\boldsymbol{P}^{(i+1,j)}=\boldsymbol{P}^{(i,j)}+\Delta\boldsymbol{P}^{(i,j)}.$
\subsubsection{Outer Iteration}
In every outer loop, a series of $N_i$ inner iterations are executed to accumulate the meta-loss. This meta-loss promotes the maximization of the SINR of the target and enforces the necessary constraints to ensure the optimization problem \eqref{optim} remains feasible. Accordingly, the meta-loss function is formulated as:
\begin{equation}
    \mathcal{L}^j= \mathcal{L}_{target}^j + \mathcal{L}_{th_s}^j + 
    \mathcal{L}_{th_d}^j +\mathcal{L}_{th_u}^j + \mathcal{L}_{up}^j + \mathcal{L}_{MA}^j, 
    \label{loss}
\end{equation}
where the term $\mathcal{L}_{target}^j$ is used to express the loss corresponding to the objective function and it is defined as:
\begin{equation}
    \mathcal{L}_{target}^j =  - \Lambda_t.
\end{equation}
\noindent The terms $\mathcal{L}_{th_s}^j,  
    \mathcal{L}_{th_d}^j, $ and $ \mathcal{L}_{th_u}^j$ allows satisfying the QoS constraints for communication and sensing \eqref{C_3} and  \eqref{C_3333}, respectively. Hence, they are defined in the following manner:
\begin{equation}
    \mathcal{L}_{C_{Q}}^j= \sum\nolimits_{k=1}^K \zeta_{C_{Q}} \lambda (Q_o), 
\end{equation}
\noindent where $C_{Q} \in \{th_s, th_d, th_u\}$, $\zeta_{C_{Q}}$ are the corresponding regularization parameters and $\lambda(Q_o)$, with $Q_o \in \{ R_{d, i}, R_{u, j}, \Lambda_t \}$, represent the indicator function expressed as:
\vspace{-0.2cm}
\begin{equation}
    \lambda  = \begin{cases}0, & \mbox{if } \mbox{$T - Q_o \leq 0$}, \\  1, & \mbox{} \mbox{otherwise}, \end{cases}
\end{equation}
\noindent where $T \in \{ R_{th,d}, R_{th,u}, \Lambda_{th,s} \}$. Now, the constraint \eqref{C_2} is satisfied using the following expression:
\begin{equation}
    \mathcal{L}_{up}^j = \sum_{j \in \mathcal{U}}   \zeta_2 I_1 (P_{u,j} ),
\end{equation}
\noindent where $\zeta_2$ represent the regularization parameter and $I_1(.)$ is an  indicator functions defined as $I_1  =0$, if $P_u^{max} \ge P_{u,j}$, and $I_1 = 1$ otherwise.
%
Similarly, the last term $\mathcal{L}_{MA}^j$ of \eqref{loss} handles the constraints \eqref{c4} - \eqref{c99} and is represented as:
\begin{equation}
    \mathcal{L}_{MA}^{j} =\sum_{n_t \in N_T} \sum_{n_{r_s} \in N_{R_c}} \sum_{n_{r_s} \in N_{R_s}}\zeta_3 I_2 (\mathcal{A}_{n_t}+\mathcal{A}_{n_{r_c}} + \mathcal{A}_{n_{r_s}}),
\end{equation}
\begin{equation}
   \mathcal{A}_{n_t}=DS-||\textbf{t}_{n_t}^{BS}-\textbf{t}_{\hat{n}_t}^{BS}|| \le 0, 1 \le n_t \neq \hat{n}_t\le N_T,
   \label{ma1}
\end{equation}
\begin{equation}
     \mathcal{A}_{n_{r_c}}=DS-||\textbf{r}_{n_{r_c}}^{BS}-\textbf{r}_{\hat{n}_{r_c}}^{BS}|| \le 0, 1 \le n_{r_c} \neq \hat{n}_{r_c}\le N_{R_c},
     \label{ma2}
\end{equation}
\begin{equation}
     \mathcal{A}_{n_{r_s}}=DS-||\textbf{r}_{n_{r_s}}^{BS}-\textbf{r}_{\hat{n}_{r_s}}^{BS}|| \le 0, 1 \le n_{r_s} \neq \hat{n}_{r_s}\le N_{R_s},
     \label{ma3}
\end{equation}
%
\noindent where $I_2 = 0$ if the expression 
\begin{equation}
\Omega \;\equiv\;
\max\!\left\{
\max_{\,n_t \neq \hat{n}_t}\mathcal{A}_{n_t},\;
\max_{\,n_{r_c} \neq \hat{n}_{r_c}}\mathcal{A}_{n_{r_c}},\;
\max_{\,n_{r_s} \neq \hat{n}_{r_s}}\mathcal{A}_{n_{r_s}}
\right\} \;\le\; 0
\notag
\end{equation}
\noindent holds and $I_2 = 1$, otherwise, and $\zeta_3$ is a regularization parameter. Now in order to guarantee the power budget constraint \eqref{C_7}, we adopt a normalization technique that is based on projecting the updated matrix $\boldsymbol{P}^{*}$ \cite{loli2024metalearningbasedoptimizationlarge}. Thus, it can be expressed as:
\begin{equation}
        \boldsymbol{P}^*  = \begin{cases}\boldsymbol{P}^*, & \mbox{if } \mbox{$\mbox{Tr}(\boldsymbol{P}^*(\boldsymbol{P}^*)^H) \leq P_{BS}$}, \\  \sqrt{\frac{P_{BS}}{\mbox{Tr}(\boldsymbol{P}^*(\boldsymbol{P}^*)^H) }}\boldsymbol{P}^*, & \mbox{otherwise}. \end{cases}
        \label{normalized}
\end{equation}
Finally, the constraints \eqref{C_5} and \eqref{tx1} were ensured by following the same technique as in \cite{khisa2025metalearningdrivenmovableantennaassistedfullduplex}, and the full details are omitted for the limited space.
\subsubsection{Epoch Iteration}
Here, the neural network's parameters are updated. This involves $N_o$ outer iterations, after which the individual losses are aggregated and their average is computed as $\bar{\mathcal{L}}=\frac{1}{N_o}\sum_{j=1}^{N_o}\mathcal{L}^j.$
In the next step, the backpropagation takes place with Adam optimizer being leveraged to update the parameters of the different sub-networks presented above. The full details of the algorithm can be found in \cite{khisa2025metalearningdrivenmovableantennaassistedfullduplex}. 

  \vspace{-0.3cm}
\section{Numerical Evaluation}

To assess the performance of the presented model, we analyze the numerical evaluation. The suggested model which is denoted as \textbf{ISAC-FD-MA} will be compared with the following benchmarks:
\begin{itemize}
  \item \textbf{ISAC-FD-MA (fmincon)}: This benchmark leverages a predefined MATLAB solver "fmincon" to solve the optimization problem of the model \cite{khisa2025metalearningdrivenmovableantennaassistedfullduplex}.
\item \textbf{FD-ISAC-MA} \textbar   \textbf{FPA-BS-R}: This model examines the FD ISAC model utilizing MAs only at the FD BS.  
\item \textbf{FD-ISAC-FPA}: This scheme utilizes FPAs at both BSs.
\end{itemize}
The model parameters adopted in this work follow \cite{khisa2025metalearningdrivenmovableantennaassistedfullduplex} and the results are averaged over 150 Monte Carlo iterations. 

In Fig. 2 (a), we examine the convergence of the presented GML approach by plotting the objective function versus the number of iterations. It is clear that with the increase in the number of iterations, the objective function keeps on increasing with fluctuations until reaching around 35.5 by the iteration 280. This signifies the ability of the model to adjust the optimization parameters to minimize the loss function. On the other hand, after the iteration 280 the objective value tends to be stable signifying the convergence of the algorithm. 

Fig. 2 (b) presents the SINR of reflected echo versus the transmit power at the FD BS. The graph shows almost a perfect match between the presented system model by adopting the GML approach and the benchmark that utilizes fmincon. This can be interpreted by the ability of the algorithm to produce almost optimal results (99\% compared to fmincon). In addition, we can observe that the SINR increases for all models with the increase in the FD BS power. That's because the BS is investing more power in the sensing functionality after satisfying the QoS requirements. Also, it is clear that the MAs play a significant role, leading to better results compared to FPAs. That can be explained by the enhanced ability to mitigate interference, battle the fading, enhance the overall performance, and establish a better tradeoff between sensing and communication functionalities.

In Fig. 2 (c), we study the effect of SI on the SINR of the received echo at BS R. The graph shows the decrease in the echo strength for all models, which can be interpreted by the need for the FD BS to further adjust the beams to satisfy the UL and DL QoS requirements. In addition, the presented system showed resilience in handling the SI due to the flexibility offered by the installed MAs.

Fig. \ref{finall} illustrates the variation of the SINR as a function of the QoS requirements for both DL and UL services. The 3D plot clearly reveals the inherent trade-off between sensing and communication, as sensing performance tends to degrade with increasing values of $R_{th,d}$ and $R_{th,u}$. This inverse relationship highlights the conflicting nature of resource utilization for both functionalities. Furthermore, the proposed model demonstrates greater robustness compared to other benchmarks, primarily due to the mobility of the MAs, which introduces additional degrees of freedom and results in fewer infeasible solutions. 

\vspace{-0.3cm}
\section{Conclusion}
In conclusion, this study demonstrates the potential of integrating MA technology within FD ISAC systems to enhance both communication and sensing performance. By jointly optimizing beamforming, power allocation, and antenna positioning, the proposed gradient-based meta-learning approach effectively addresses the complexity of the formulated non-convex optimization problem. Simulation results validate that the MA-assisted scheme not only achieves near-optimal performance (99\%) but also surpasses traditional FPAs benchmarks.
\bibliographystyle{IEEEtran}
  \vspace{-0.7cm}
\bibliography{ref}

\end{document}